\documentclass[runningheads]{llncs}
\usepackage[T1]{fontenc}
\usepackage{graphicx}
\usepackage{caption}
\usepackage{enumitem} 
\usepackage{csquotes} 
\usepackage{xcolor}  
\renewcommand{\enquote}[1]{\textit{``#1''}}
\usepackage{comment}
\usepackage{array} 
\usepackage{setspace} 

\usepackage[bookmarks=true,bookmarksnumbered=true,colorlinks=true,linkcolor=blue,urlcolor=blue]{hyperref}

\begin{document}
\title{From Earthquake Solidarity to Educational Equity: Conceptualizing a Sustainable, Volunteer-Driven P2P Learning Ecosystem at Scale}
\titlerunning{From Earthquake Solidarity to Educational Equity}
%
\author{Öykü Kaplan\inst{1} \and
Adam Przybyłek\inst{1,2}\orcidID{0000-0002-8231-709X} \and
Michael Neumann\inst{3}\orcidID{0000-0002-4220-9641} \and
Netta Iivari\inst{4}\orcidID{0000-0002-7420-2890}
}
\authorrunning{Ö. Kaplan et al.}
%
\institute{Gdańsk University of Technology, Poland
\email{oykukaplan01@gmail.com} \and
University of Galway, Ireland
\email{adam.przybylek@gmail.com} \and
University of Applied Sciences and Arts Hannover, Germany
\email{michael.neumann@hs-hannover.de} \and
University of Oulu, Finland
\email{Netta.Iivari@oulu.fi}
}
\maketitle              
\begin{abstract}
This study examines the evolution of a grassroots, volunteer-driven peer-to-peer (P2P) educational initiative from an emergency response to the 2023 Türkiye earthquake into a sustainable ecosystem that operated for over two years and supported 300+ middle-school learners with 40+ volunteer tutors. Employing an interpretive case study approach, we triangulated data from participant observation, focus groups, questionnaires, and collaborative visioning workshops to investigate the socio-technical dynamics enabling long-term resilience in a fully online, nonreciprocal far-peer tutoring setting. Our findings reveal that while age proximity fosters trust and open communication, it also poses challenges for tutors who must balance peer rapport with instructional authority. Volunteer engagement is driven primarily by intrinsic motives—educational impact and community belonging—while optional micro-earning is envisioned as a practical enabler for long-term sustainability. Tutees report significant gains in confidence, self-expression, and accelerated comprehension, attributing these outcomes to personalized, interactive sessions within a ``family-like'' safe space that combines academic instruction with socio-emotional support. Notably, tutees view tutors as aspirational role models and express strong intentions to return as tutors themselves, envisioning a self-regenerating cycle of intergenerational reciprocity that  carries knowledge and solidarity from generation to generation. Both cohorts call for a dedicated platform featuring integrated scheduling, personalization, feedback, and quality assurance mechanisms. We synthesize these insights into theory-informed implications and five design principles for sustainable P2P learning ecosystems at scale.

\end{abstract}
\section{Introduction} 
Volunteer-driven tutoring has gained prominence worldwide as a strategy to support learners, particularly in response to crises or to address educational inequities~\cite{Hardyman_etal_2020,Child_2020}. In such contexts, peer-assisted learning (PAL) strategies can extend educational opportunities, especially to those excluded from formal systems~\cite{Capp_etal_2018}. Grounded in cognitive and social constructivism, peer-to-peer (P2P) learning leverages social interactions to foster cognitive development, creating collaborative environments where knowledge is co-constructed through peer engagement ~\cite{Rasheed_etal_2021,Mendieta_etal_2023,Kucharska_etal_2024,Hidayat_Saad_2025}. Research highlights that P2P learning enhances academic achievement, social connectedness, and psychological well-being~\cite{Tibingana_etal_2022}, proving especially beneficial for students with limited cultural, social, or academic capital~\cite{Donald_Ford_2023}. Moreover, PAL benefits both tutors and tutees~\cite{Rasheed_etal_2021,Tibingana_etal_2022,Byl2023,gutierrez2024,Hidayat_Saad_2025}. Further refining the concept, Byl \& Topping \cite{Byl2023} distinguish between peer tutoring (focused on subject knowledge) and peer mentoring (emphasizing supportive relationships). They also differentiate reciprocal peer feedback (two-way, between similar near-peers) from nonreciprocal feedback (one-way, between far-peers with differing age or skill levels).

Advances in Information and Communication Technology (ICT) and expanding Internet access have made digitally mediated collaboration increasingly feasible, lowering the barriers to organizing online P2P learning beyond formal classroom settings. The COVID-19 pandemic further accelerated the shift to online learning~\cite{Ng_Przybylek_2021,Neumann.2021,Neumann.2022,Mendieta_etal_2023,Marcinkowski_2020}, compelling educational institutions worldwide to rapidly adopt tools like Zoom and Microsoft Teams on an unprecedented scale to ensure educational continuity~\cite{He_etal_2023,Sharlovych_etal_2023}. Within this digital landscape, online learning communities emerged as indispensable support structures, connecting individuals to collaboratively exchange knowledge and expertise independent of formal institutions~\cite{Zamiri_Esmaeili_2024}.

The utility of such decentralized models extends beyond pandemics to other crises—including natural disasters and military conflicts—where conventional education systems falter due to physical infrastructure damage, displacement, and emotional distress~\cite{Edtech_2024}. This potential materialized in March 2023 with the launch of the Earthquake Solidarity Project, a grassroots effort established to deliver both academic and emotional support to middle school students in southern Türkiye following a devastating natural disaster. This study reports on the evolution of that initiative into a two-year-plus educational program and, to our knowledge, offers the first empirical study of volunteer-driven, nonreciprocal far-peer tutoring and mentoring delivered entirely online in a post-disaster setting.

This article constitutes a revised and extended version of our ISD'25 paper~\cite{Kaplan_2025}. While the original study focused on a retrospective analysis of the Earthquake Solidarity Project, this extension incorporates findings from collaborative visioning workshops to explore the ecosystem's future trajectory, scalability, and long-term sustainability. Informed by this real-world case, our research distills transferable principles for designing and scaling similar P2P initiatives, thereby helping to address educational inequities and enhance educational access for students in need more broadly. This work directly responds to the conference theme—\enquote{Empowering the Interdisciplinary Role of ISD in Addressing Contemporary Issues in Digital Transformation}—by demonstrating how a socio-technical approach can harness digital affordances to transform latent volunteer capacity into a resilient learning ecosystem.

\noindent
Our investigation was guided by three research questions:
\begin{enumerate}[
    label=\textbf{RQ\arabic*:},
    leftmargin=1.2cm,
    labelsep=0.4em,
    itemsep=4pt,
    topsep=6pt
]
\item[RQ1:] What factors influence the motivation and sustained participation of volunteer tutors and learners in online P2P educational programs?
\item[RQ2:] How do tutors and learners perceive the effectiveness of the online P2P learning model?
\item[RQ3:] What barriers, facilitators, and key design principles underpin the scaling of volunteer-driven, online P2P educational ecosystems toward widespread adoption?
\end{enumerate}

The rest of this paper is organized as follows. Section~\ref{sec:literature} reviews prior work on peer-assisted learning and volunteer-driven, community-based education, highlighting the research gap this study addresses. Section~\ref{sec:methodology} details the research methodology, outlining the case context, data collection techniques, and analytical approach. Section~\ref{sec:findings} reports the findings derived from documentation, fieldnotes, focus groups, questionnaires, and visioning workshops. Section~\ref{sec:discussion} interprets these results through established theoretical lenses and derives theoretical implications and design principles for sustainable P2P learning ecosystems. Finally, Section~\ref{sec:conclusion} concludes the paper and outlines avenues for future research.

\section{Literature Review}
\label{sec:literature}
\subsection{Peer-Assisted Learning}
A large body of research demonstrates the effectiveness of peer tutoring programs in supporting students across multiple domains \cite{Capp_etal_2018,Hidayat_Saad_2025}. These programs promote not only academic success but also positive social and emotional outcomes \cite{Capp_etal_2018,Hidayat_Saad_2025}. In a meta-analysis of 24 studies, Hidayat \& Saad~\cite{Hidayat_Saad_2025} found that peer tutoring significantly improves academic performance and yields modest but meaningful gains in students' self-concept, critical thinking, and attitudes toward school. Furthermore, Donald \& Ford~\cite{Donald_Ford_2023} suggest that peer learning extends beyond academic development by connecting students to essential support services and fostering a sense of belonging through collaborative peer networks.

Beyond overall effectiveness, research has examined the specific structures of these interactions, particularly comparing nonreciprocal models with reciprocal models. Studies focusing on nonreciprocal arrangements show consistent benefits across different age groups. For instance, in primary education, Tymms et al. \cite{Tymms_etal_2011} found that nonreciprocal peer tutoring, involving older tutors (aged 10) and younger tutees (aged 8), produces modest but consistent gains in reading and mathematics attainment for both groups. Similarly, Capp et al. \cite{Capp_etal_2018} reported that a nonreciprocal peer tutoring program in a K–8 school, with middle-school students tutoring upper-elementary tutees, led to modest academic gains but more substantial social–emotional benefits, high satisfaction among participants and parents, and practical recommendations for future implementation. Building on this body of research, Byl \& Topping \cite{Byl2023} conducted a more nuanced investigation within a university setting that not only examined the reciprocal versus nonreciprocal dimension but also distinguished between peer tutoring and peer mentoring. Their study of first-year students found that different nonreciprocal models yielded distinct advantages: nonreciprocal peer tutoring delivered the greatest gains in academic engagement and involvement, while nonreciprocal peer mentoring proved most effective in enhancing both social integration and persistence. The particular success of nonreciprocal peer mentoring in boosting student persistence was attributed to the unique position of the mentors — typically higher-year students who had navigated the same early university stage relatively recently. This proximity in experience allowed mentors to better relate to their mentees' challenges, offer relevant guidance based on their greater experience and understanding of common struggles, and serve as effective role models \cite{Byl2023}. 

Despite the demonstrated benefits of PAL, its potential has been less explored and realized in online and hybrid educational settings. Research highlights a critical gap between the potential of digital platforms to facilitate PAL and the actual engagement levels achieved in practice. Kotturi et al. \cite{Kotturi_etal_2015} emphasize that merely providing technological tools for peer interaction is insufficient; without explicit curricular integration and performance incentives, student participation often falters, as learners fail to perceive the inherent value in these collaborative activities. Rasheed et al. \cite{Rasheed_etal_2021} addressed this by scaffolding online PAL through structured group formation, system design features fostering collaboration, and incentives against social loafing. Their experiment with 120 students showed significant academic gains when participants were prepared and motivated to engage actively. Building on this foundation, Mendieta-Aragón et al. \cite{Mendieta_etal_2023} found that students acting as asynchronous video tutors in hybrid courses reported enhanced motivation, mastery, creativity, and communication skills. Tutees likewise preferred peer-generated videos over traditional materials. Synthesizing these studies, three key lessons emerge: (1) structured facilitation is crucial in digital PAL ecosystems; (2) incentives (grading, badges, recognition) counteract participation fatigue; and (3) reciprocal value creation (e.g., alternating tutor/tutee roles) deepens engagement.

However, these insights are drawn from stable academic environments, leaving their applicability to post-crisis settings—characterized by unique challenges of scalability, volunteer retention, and learner resilience—largely unexplored.

\subsection{Volunteer-Driven and Community-Based Education}
The scalability of PAL hinges on volunteerism and community engagement. Large-scale initiatives, such as Action Tutoring in the UK, show that programs thrive when fueled by intrinsically motivated volunteers seeking social impact, yielding academic gains for pupils and significant personal benefits for the volunteers themselves \cite{Hardyman_etal_2020}. The importance of community ownership and local leadership is further magnified in unstable or resource-scarce contexts. Studies in post-conflict and rural settings demonstrate that sustainability is built on genuine community participation, trust in local leadership, and flexible program structures tailored to community needs \cite{Taniguchi_Hirakawa_2016,Faizi_2017}. In essence, sustainable programs are those co-created and actively supported by the community. The convergence of volunteerism, community engagement, and technology becomes particularly relevant in post-disaster recovery. Reporting on a service-learning initiative after Hurricane Katrina, Evans-Cowley \cite{Evans_2006} described how volunteer professionals guided students in disaster-recovery projects, using online forums and project-management tools to sustain collaboration across geographic distances. Collectively, these studies highlight core principles of motivation, ownership, and flexibility, but leave a critical gap in understanding their application in fully online settings.

\subsection{Research Gap}
While substantial research has established the benefits of PAL and explored the dynamics of volunteer-driven community education, these domains have largely been investigated separately, with limited attention to their integration or the specific affordances of educational technology. The literature highlights successful models in stable academic environments or community-based initiatives, but a critical gap emerges at their intersection, particularly in volatile settings.

This gap is particularly evident in post-disaster contexts, where challenges of compromised infrastructure, inconsistent access, and heightened emotional needs demand flexible socio-technical systems. To our knowledge, no empirical study has comprehensively investigated a volunteer-based, nonreciprocal far-peer tutoring and mentoring model delivered entirely online in such a setting.

By addressing this multifaceted gap, this study moves beyond documenting a single case to offer transferable principles for designing a scalable and participatory online P2P learning ecosystem that integrates motivated volunteers, purpose-built digital platforms, and diversified funding mechanisms.

\section{Research Methodology} 
\label{sec:methodology}
We employed an interpretive case study methodology to uncover meaning through participant interpretations of their context \cite{Klein_Myers_1999}. Aligned with our exploratory goals \cite{Robson_2002}, this approach facilitated a holistic examination of volunteer motivations, learner engagement, the operational intricacies of the online P2P ecosystem, and the broader contextual factors that enabled its sustainability beyond the initial crisis.

The Earthquake Solidarity Project was selected as a single holistic case \cite{Yin_2009} due to its unique trajectory from emergency response to a two-year-plus educational program \cite{Benbasat_etal_1987}. Through this case, we demonstrate how this solidarity-based approach not only addressed urgent needs but also fostered a resilient, community-centric solution for enduring educational support.

\subsection{Case Description}
The Earthquake Solidarity Project was launched in March 2023 in response to the devastating earthquake that hit southern Türkiye on 6 February. The initiative's founder, driven by firsthand experiences with past earthquakes, established the program to offer both academic and emotional support to affected middle school students. The post-disaster educational landscape was dire: schools were destroyed or damaged, reconstruction was slow, and many students lacked access to safe school buildings for months.

Across the affected regions, traditional, in-person classes were suspended. In Hatay, one of the hardest-hit provinces, schooling resumed incrementally over subsequent months, with existing facilities shared among multiple schools in a double-shift system. Compounding these challenges, widespread family relocation due to housing destruction severed students' connections to their teachers and peers, fundamentally disrupting educational continuity.

The program targeted students in grades 6 to 8, with mathematics chosen as the core subject due to its foundational role in the curriculum and the founder’s expertise. Volunteer tutors were recruited through online forms and brief interviews; tutees applied via the same process. The volunteer tutors came from diverse academic backgrounds — including undergraduate, master's, and PhD students, as well as recent graduates — and participated remotely from various Turkish cities. Before the program began, all volunteers attended an online orientation on trauma-informed pedagogy, led by a licensed psychologist. A flexible 12-week curriculum based on the national syllabus was crafted and tailored to the needs of each class. Although the project began without institutional funding, it later secured modest sponsorships to cover essential operational costs such as Zoom subscriptions and digital learning materials. To promote personalized support, the program utilized small classes of 5–6 students per tutor. The first author, along with three members of the Hypatia community, coordinated scheduling, class assignments, and weekly follow-ups. Beyond academic instruction, the initiative also aimed to foster emotional connection, a sense of belonging, and solidarity. 

Initially designed as a short-term intervention ending in June 2023, the project’s success and the strong commitment of both tutors and learners led to its extension until all enrolled students completed middle school in June 2025. In total, the initiative supported over 300 learners and engaged more than 40 volunteer educators, demonstrating remarkable sustainability and growth beyond its original expectations.

\subsection{Data Collection Techniques}
To gain a holistic understanding of the Earthquake Solidarity Project, we employed multiple data collection techniques, enabling methodological triangulation and the capture of rich qualitative and quantitative insights. All collected data are available at \url{https://p2p.lasd.pl}.

\paragraph{Documentation Analysis.}
We began by analyzing project documentation, specifically the Tutoring Session Report—a concise, regularly updated spreadsheet in which volunteer tutors recorded session dates, tutee attendance, brief lesson synopses, and overall session status. We also examined the first author's diary, which provided rich, firsthand accounts of the project's inception and early operation. These diary entries offered invaluable insights into initial motivations, challenges, and the project's evolution from an insider's perspective. Together, these documents serve as critical data illuminating the project's formative context and operational dynamics, bridging the temporal gap between the project's launch and the commencement of our formal research. 

\paragraph{Participant Observation.}
Once the research commenced, the first author adopted the role of a participant observer. In line with Spradley \cite{Spradley_1980}, she engaged in ``complete participation'', transforming her ordinary involvement into a deliberate ethnographic stance. At this stage, the most essential diary entries from the pre-research period were systematically reviewed and converted into structured fieldnotes to ensure methodological consistency. From that point forward, she continued diligently recording fieldnotes during significant events and interactions, thereby capturing key aspects of the project's ongoing evolution and operational dynamics as they naturally unfolded.

\paragraph{Focus Groups.}
As the project neared completion, we held four semi-structured focus group sessions with key stakeholders: two with learners (L1: n=5, 58 minutes; L2: n=5, 50 minutes) and two with volunteer tutors (T1: n=4, 54 minutes; T2: n=5, 70 minutes). We employed purposive sampling to select learners who represented a variety of age groups (ranging from 13 to 15 years old) and who had interacted with different tutors during the program, enabling us to capture a wide range of experiences shaped by diverse teaching styles and interpersonal dynamics. Four participants had completed the program one or two years prior to the focus group, providing sufficient temporal distance to reflect holistically on the initiative's long-term impact. Tutors were similarly chosen for their diversity in educational and personal backgrounds, and gender, with all having experience working with learners from different age groups while being at various life stages—including recent graduates, master's students, and early-career professionals—with ages spanning 22 to 29. This heterogeneity offered a richer view of motivations, challenges, and pedagogical strategies.

Guided by a set of open-ended questions, the sessions explored participant experiences, challenges, and perspectives on the P2P education model. Tutor groups also discussed strategies for scaling and sustaining similar volunteer-driven P2P education initiatives, with an emphasis on volunteer motivation and willingness to participate. All participants consented to the use of their real names.

\paragraph{Questionnaires.}
To complement our qualitative findings and reach a wider set of participants, we administered two structured questionnaires—one for tutees and another for tutors. Both instruments primarily used 5-point Likert-type items, supplemented by an open-ended question to elicit additional feedback. The tutee questionnaire focused on emotional support, academic improvements, satisfaction with the P2P model, and learning format preferences. The tutor questionnaire explored preparedness, perceptions of program effectiveness, key motivators, and commitment to future volunteering. Notably, the tutee response rate was relatively low, likely because outreach relied on the project's original communication channel, which many graduates from 2023 and 2024 had stopped actively monitoring.

\paragraph{Visioning Workshops.}
To explore the future trajectory of the ecosystem, we conducted two online co-design workshops using the \emph{Cover Story} technique~\cite{Zakrzewski_2018}—one with tutors (n=5) and one with tutees (n=5). \emph{Cover Story} is a collaborative visioning game from the broader genre of serious games that have recently gained traction in software engineering research~\cite{Olszewski_2016,Wawryk_Ng_2019,Ng_2020,Mich_Ng_2020,Ng_Kuduk_2024,Ng_2024,Ng_2025}. It facilitates expansive thinking about a desired future state, surfacing aspirations, values, and latent challenges that may not emerge through conventional interviews~\cite{Zakrzewski_2018}. In our sessions, participants imagined an ideal P2P educational platform so spectacular that it would be featured on the front page of a magazine, and then collaboratively designed this fictional cover. Each session concluded with a moderated discussion to deconstruct the design choices, yielding deep insights into stakeholders' visions for scalability, sustainability, and the platform's evolving identity. Screenshots of the designed covers are available in our online repository.

\subsection{Data Analysis Techniques}
To derive meaningful insights from the diverse data collected, we employed both qualitative and quantitative analysis techniques tailored to the specific characteristics of each data source. Given the volume and heterogeneity of the data, our analytical process began with independent examinations, where each dataset was scrutinized in isolation using the most appropriate method. Only after distilling preliminary insights from each source did we undertake systematic triangulation across data sources and analytical techniques. This approach enabled us to construct robust interpretations anchored in multiple lines of evidence, thereby enhancing the validity and depth of our findings.

\paragraph{Thematic Analysis.}
Thematic analysis \cite{Braun_Clarke_2006} served as the cornerstone for interpreting qualitative data from four sources: fieldnotes, focus groups, open-ended questionnaire responses, and visioning workshops. We selected this method for its flexibility in identifying, analyzing, and reporting patterns (themes) within qualitative data, making it well suited for capturing the rich narrative insights these sources provided. To ensure a meticulous, context-sensitive analysis, we conducted independent thematic analyses for the tutor and tutee datasets, allowing us to immerse ourselves fully in each stakeholder perspective without risking the conflation of narratives from distinct participant groups. The resulting codebooks, defining the codes and categories, are available at \url{https://p2p.lasd.pl} to ensure transparency and replicability.


\paragraph{Diverging Stacked Bar Charts and Descriptive Statistical Analysis.} We visualized responses to Likert-type items using diverging stacked bar charts. This technique provided an intuitive graphical summary of participants' attitudes, levels of satisfaction, and perceptions of effectiveness while respecting the ordinal nature of the data. Additionally, we employed descriptive statistics to summarize other quantitative measures, such as volunteers' intended time commitments for future initiatives. Because our questionnaires targeted the entire program population, inferential statistics were not applicable.

\subsection{Threats to Validity} 
To systematically address potential limitations, we structure our discussion of threats to validity following established frameworks for case studies \cite{Yin_2009,Runeson_Host_2009}. However, as our study is exploratory and does not seek to establish causal relationships, internal validity is not investigated.

\paragraph{Construct Validity.}
A potential threat to construct validity is that the focus group sessions were moderated by the project founder. Specifically, recipients of free tutoring might have felt obliged to provide positive responses due to social desirability bias stemming from gratitude for the program. To reduce this bias, we emphasized the value of critical feedback during sessions and offered an anonymous questionnaire channel. Another threat lies in the risk of misinterpreting what participants said or over-relying on subjective data (the reliance on participants' perceptions, feelings, experiences, and memories), which may reflect biased or incomplete perspectives. To mitigate both these threats, we employed data and method triangulation.

\paragraph{External Validity.}
The Earthquake Solidarity Project is a unique and highly context-specific case. Therefore, direct generalization of its findings to other educational initiatives, particularly those operating in different cultural contexts or without a comparable crisis catalyst, is limited. Nonetheless, our study provides rich, transferable insights into the P2P education model. These insights can theoretically inform and practically guide similar community-driven projects.

\paragraph{Reliability.}
A potential threat to reliability in this study stems from the inherently interpretive nature of qualitative data analysis, particularly in thematic analyses, where different researchers might derive varying insights from the same dataset.

Moreover, the first author's role as a participant-observer introduces a unique contextual lens that, while enriching the data with insider perspectives, may limit replicability by others unfamiliar with the project's history. To mitigate this, we ensured that fieldnotes were systematically recorded and cross-referenced with other data sources. We also reduced the risk of misinterpretation by audio-recording and transcribing verbatim all focus group sessions. 

To enhance reliability, two researchers independently analyzed the data. Their findings were then systematically compared and reconciled through discussion to resolve discrepancies, achieve consensus, and enrich the final interpretations.

Finally, we documented our data analysis procedures in detail, providing a clear audit trail of our research process. For instance, the codebooks developed during thematic analysis, along with the coded transcriptions, are made available to enhance the traceability of our interpretations. 

\section{Findings}
\label{sec:findings}
\subsection{Insights from Documentation and Participant Observation}
Analysis of tutoring session reports from the project's initial phase, a 12-week program concluding at the end of the Spring 2023 semester, reveals key operational dynamics within the digitally-mediated, post-disaster context. Persistent technical challenges, particularly internet connectivity issues for students in rural or temporary housing, were recurring impediments to consistent attendance, yet overall engagement remained high. Tutors demonstrated adaptability and commitment, proactively delivering one-on-one follow-up sessions for students with repeated absences. In cases of student attrition, follow-ups with families identified geographic relocation and persistent connectivity barriers as primary causes. Moreover, tutors responded dynamically to evolving student needs; noting learner fatigue in the final three weeks, they integrated more interactive and game-based activities to sustain motivation. From Week 8 onward, scheduling conflicts also emerged as a notable attendance constraint.

The program’s socio-emotional impact was evident at the voluntary end-of-semester virtual graduation ceremony. Attended by over 100 participants, this event showcased the project's broader impact through powerful emotional testimonials from students and parents. One student vividly captured the sentiment, stating, \enquote{Everything was dark... You became our stars and illuminated our path.} This culminating event highlighted the strong community spirit and significant emotional support fostered by the initiative, extending its value beyond academic outcomes.

In the subsequent 2023/2024 academic year, the initiative continued with a reduced cohort of returning students (7th and 8th graders). The program expanded beyond core academics, offering specialized activities such as peer mentorship sessions where successful high school students shared study strategies and transitional experiences. This period of sustained success, set against the clear limitations of Zoom, laid the groundwork for a more ambitious solution: a dedicated P2P platform.

During the 2024/2025 academic year, the program entered its final phase, exclusively supporting 8th grade students as they completed middle school. This further reduced student-to-tutor ratio facilitated enhanced personalization and responsiveness, reinforcing strong tutor–student bonds and maintaining high motivational levels among participants.

Participant observation fieldnotes provide additional insights, capturing the enduring nature of the project's impact long after the initial crisis response.  For example, two years after its inception, on the anniversary of the earthquake, a parent shared: \enquote{You were there when no one else held our children's hands. My older daughter graduated with your help, and my younger one is still learning with you. This meant everything to us.}

Analysis of additional participant observation fieldnotes revealed the potential for scaling the initiative into a nationwide program to support students in need. At the same time, these notes underscore the inherent limitations of the current approach, including constrained volunteer availability, lack of financial support, and the inadequacy of generic platforms like Zoom for fostering deep interaction and community. It was this very tension—between the project's demonstrated positive impact and its underlying vulnerabilities—that catalyzed a vision for a dedicated P2P education platform. Such a platform would establish a sustainable ecosystem for peer-driven learning, addressing educational inequities across Turkey through dedicated features that support meaningful connections and growth for both learners and university-level tutors.

\subsection{Focus Group Insights: Volunteer Tutor Perspectives} 
\subsubsection{Operational Challenges and Pedagogical Considerations in P2P Tutoring}
\paragraph{Technical and Environmental Barriers.}
Tutors recognized that online P2P learning both broadens access and introduces new challenges. While digital lessons enable learners in remote or underserved areas to receive quality support, sessions are sometimes cut short by internet dropouts, power outages, or a lack of appropriate devices. Tutors also pointed out in-home distractions—such as family interruptions—that undermine learners’ concentration.

\paragraph{Interaction and Engagement Challenges in Online Learning.} Tutors agreed that passive lectures fail to capture attention, and that monitoring engagement online is considerably harder than in a traditional classroom. To address this, they  recommended incorporating interactive elements—breakout discussions, real-time quizzes, and collaborative annotation tools—alongside socio-emotional check-ins to recreate the supportive atmosphere of in-person classes. When these features were effectively integrated, tutors observed notably higher learner motivation and deeper cognitive involvement.

\paragraph{The Double-Edged Sword of Age Proximity: Fostering Trust and Challenging Authority.}
Teaching peers close in age was identified as both an opportunity and a challenge. Tutors agreed that age proximity lowers affective barriers, fostering trust and openness, and that near-peer tutors can better understand tutee challenges from their own recent experience. However, this proximity also creates authority gaps: several tutors feared that tutees might view them as ``just another friend,'' making classroom management more difficult.

\paragraph{Personalization and Learner-Centered Approaches.}
Volunteers recognized that learning preferences vary widely: some tutees thrive in groups, others in one-on-one settings. As Nefise (T2) observed, \enquote{In a group setting, not everyone is comfortable being open. If a tutee doesn't understand something, they might hesitate to ask.} Accordingly, an effective P2P platform should offer both formats to accommodate these varied needs.

Personalization emerged as critical for digital P2P effectiveness. Participants consistently emphasized that tailoring lessons to individual preferences substantially improves outcomes. Elif Naz (T2) recommended that \enquote{a short test should be done before matching a mentor and a tutee. Personality compatibility is crucial.} This focus on personalization highlights a key advantage of P2P models over more standardized educational approaches.

\subsubsection{Systemic Enablers and Constraints for Volunteer-Driven P2P Initiatives}
\paragraph{Flexible Time Commitment and Micro-Volunteering.}
Time availability emerged as the key constraint for volunteering in future initiatives. Although most tutors reported that committing 1 to 2 hours per week was realistic, they also noted that their capacity could expand during emergency periods, suggesting that perceived urgency could boost volunteer engagement. They consistently stressed that academic and personal obligations fluctuate, making fixed schedules challenging. Consequently, they advocated for micro-volunteering options, proposing that instead of rigid weekly commitments, allowing tutors to select sessions on an ad-hoc basis would better align learner needs with tutor availability.

\paragraph{Multifaceted Motivation Drivers for Volunteer Tutors.}
Our analysis uncovered a rich motivation framework driving tutor engagement, comprising both intrinsic and extrinsic factors. At the core, creating educational impact consistently ranked highest as an intrinsic motivator. As Elif (T1) reflected, \enquote{Being able to contribute to the younger generation both academically and socially would be deeply meaningful.} This sense of purpose was further reinforced by the emotional fulfillment of witnessing tutee growth. Belonging to a supportive community represented another strong intrinsic driver, with Elif Naz (T2) noting that \enquote{being part of something that outlives its individual members is very meaningful.}

On the extrinsic side, professional development benefits — such as enhancing one’s CV and receiving formal recognition (reference letters, certificates) — were frequently cited by career-focused volunteers. While financial compensation was mentioned by several participants, it was consistently positioned as supplementary rather than primary.

\paragraph{Multi-Modal Feedback Channels.}  
Tutors expressed varied preferences for feedback, yet there was strong consensus on the need for anonymous mechanisms. Şebnem (T2) noted, \enquote{anonymity helps ensure more transparent and sincere feedback,} and Elif (T1) added, \enquote{Sometimes students can't express themselves face-to-face, but in written and anonymous formats, they may be more open.} Several tutors also valued direct one-on-one discussions to clarify specific points.

\subsection{Focus Group Insights: Learner Perspectives}
\subsubsection{The Power of Age Proximity and Empathetic Tutoring}
Middle school learners consistently highlighted the value of reduced hierarchical distance between themselves and their tutors. They emphasized that a smaller age gap facilitates more comfortable and effective communication. As Umut (L2) explained, \enquote{teachers who are close to our age understand us more easily.} Participants also reported that the tutors could more effectively convey complex concepts compared to their regular teachers, attributing this to generational proximity and similar thought patterns.

When discussing their ideal interaction styles, learners expressed a preference for a balanced mix of guidance and camaraderie. As Elif Dağ (L2) articulated, \enquote{It would be best if the communication is both guidance-focused and friendly—someone who gives direction while also maintaining a warm and approachable tone.}

\subsubsection{Empowering Confidence and Self-Expression}
Learners highlighted the positive impact of peer tutoring on their confidence and self-expression. As Rüveyda (L1) shared, \enquote{One of the most important things for me when I joined these sessions was how much it boosted my self-confidence. I started speaking better, expressing myself more clearly, and improving my social skills.}

This transformation extended beyond the program itself, transferring to regular school settings. Hicran (L2) contrasted  the two environments: \enquote{In a classroom setting, if there's something I don't understand, I don't speak up… But here, I can say it easily.} Supporting this observation, Azra (L2) confirmed, \enquote{Communicating with university students \{tutors\} had an impact on me—I started expressing myself more confidently at school.} 
Beyond confidence gains, learners reported cognitive benefits as well. Hicran (L2) noted accelerated comprehension: \enquote{Something that would take me three weeks to grasp on my own, I can understand here in one week.}

\subsubsection{Benefits of Small Groups and Personalized Interaction}
Learners described the P2P learning environment as one that fosters a more personalized and interactive experience. They frequently cited smaller class sizes as a key factor enabling greater participation and individualized attention. As Elif Dağ (L2) noted, \enquote{Smaller groups are better. Because the tutors are close to our age, we feel more comfortable. They motivate us.}

Learners also found this setup more conducive to concentration and engagement. They highlighted the benefit of being able to revisit challenging concepts and receive immediate feedback. Melek (L2) identified this as \enquote{the biggest advantage,} explaining that tutees get \enquote{a second chance to hear something we didn't understand at school—or learning it here first and reinforcing it later in school.}

\subsubsection{Technical and Logistical Barriers}
The online P2P model presented several practical obstacles. Connectivity issues and home distractions interrupted the learning flow. Azra (L1) cited \enquote{power outages or no internet,} while Hicran (L2)  raised concerns about the health effects of \enquote{extended time spent in front of a screen.} 
The home learning environment itself could also present challenges. Umut (L2) remarked, \enquote{Sometimes we have guests at home, and it becomes hard to find a quiet space. That makes it harder to concentrate and stay focused.} Scheduling also posed a persistent challenge, with Başak (L1) identifying timetable conflicts as \enquote{the biggest issue} and Elif Dağ (L2) noting, \enquote{lesson times sometimes clash with other plans.}

\subsubsection{Envisioned P2P Learning Platform Features}
When asked to envision an ideal P2P platform, learners identified several key features. They emphasized the value of flexibility in session formats, supporting both one-on-one sessions (for depth and personalized pacing) and small-group formats (for peer motivation and collaborative problem-solving). To manage content effectively, they suggested a shared digital library containing practice exams and guided walkthroughs, complemented by occasional ``inspiration sessions'' where tutors could share study strategies and success stories. They also called for administrative tools such as a built-in scheduling system to streamline session coordination.

To track progress effectively, participants proposed a three-tiered approach: (1) automated self-assessments and mock exams feeding into an AI-powered dashboard that pinpoints recurring errors and recommends personalized practice plans; (2) qualitative tutor feedback providing concise comments on strengths and areas for growth; and (3) anonymous leaderboards to foster healthy competition and sustain motivation. Considering broader implications, learners envisioned that a dedicated P2P platform, if implemented at scale, could democratize educational access, reduce reliance on private tutoring centers, and cultivate global learning communities.

\subsection{Questionnaire Insights} 
Quantitative data confirms the overwhelmingly positive experiences of both volunteer tutors and learners (see Figure~\ref{questionnaires}). While Zoom was viewed as an adequate tool for P2P sessions, both groups strongly advocated for a dedicated educational platform equipped with features like dedicated assignment areas, student discussion forums, and enhanced whiteboard functionality.

\hspace*{-2.0cm} 
\begin{tabular}{m{0.55\textwidth} @{\hspace{1.0cm}} m{0.55\textwidth}}
  \makebox[\linewidth][c]{\textbf{Volunteer Teacher Questionnaire (n=31)}} & \makebox[\linewidth][c]{\textbf{Tutoring Receiver Questionnaire (n=45)}} \\  
  \includegraphics[width=\linewidth]{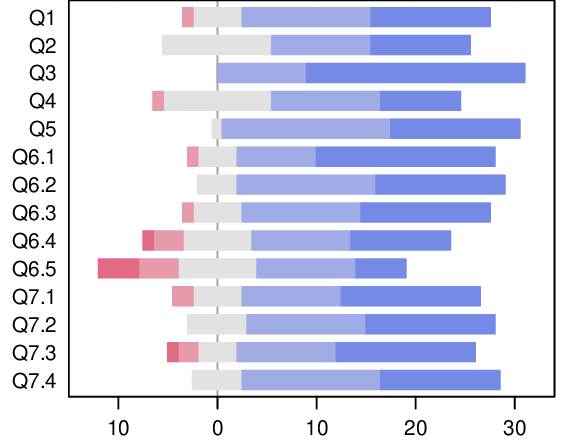} & 
  \includegraphics[width=\linewidth]{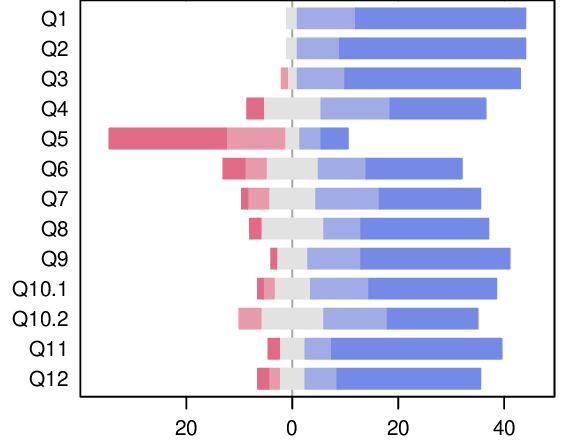} \\  
  \begin{minipage}[t]{\linewidth}  
\scriptsize 

\textbf{Reflections on Your Experience in the Earthquake Solidarity Project}

\begin{enumerate}[label=\arabic*., nosep, itemsep=2pt, topsep=4pt, left=0pt]
    \item I felt adequately prepared to teach students using the P2P model.
    \item One-on-one tutoring is more effective than group sessions for student success.
    \item Personalized support like individual mentorship greatly enhances student learning outcomes.
    \item Zoom as the primary tool has been sufficient for conducting effective P2P sessions.
    \item A dedicated P2P educational platform would significantly improve my ability to deliver lessons effectively.
\end{enumerate}

\vspace{0.5em}
\textbf{Perspectives on Future Peer-to-Peer Education Initiatives}

\begin{enumerate}[label=\arabic*., nosep, itemsep=2pt, topsep=4pt, left=0pt]
    \setcounter{enumi}{5}
    \item The following rewarding aspects may motivate me to provide lessons to students in need:
    \begin{enumerate}[label*=\arabic*., nosep, itemsep=2pt, topsep=2pt, left=1em]
        \item Social Impact
        \item Professional Development
        \item Community Building
        \item Professional Recognition (e.g., reference letters, certificates)
        \item Financial Compensation
    \end{enumerate}
    \item I would like to receive feedback via the following methods:
    \begin{enumerate}[label*=\arabic*., nosep, itemsep=2pt, topsep=2pt, left=1em]
        \item Anonymous Surveys
        \item Feedback Boxes
        \item One-on-One Discussions
        \item Analytical Reports based on data collected by the platform
    \end{enumerate}
\end{enumerate}

\end{minipage}
&
\begin{minipage}[t]{\linewidth}
\scriptsize

\textbf{Reflections on Your Experience in the Earthquake Solidarity Project}
\begin{enumerate}[label=\arabic*., nosep, itemsep=2pt, topsep=4pt, left=0pt]
    \item Participating in the tutoring program provided me with emotional support and a sense of stability while coping with the aftermath of the earthquake.
    \item This project has increased my motivation to continue my education.
    \item The free tutoring sessions have greatly improved my understanding of the subject matter.
    \item I prefer learning from university student volunteers over regular classroom teachers.
    \item I am concerned about the credibility and accuracy of the information shared by volunteers.
    \item I prefer one-on-one tutoring sessions over group lessons.
    \item Using Zoom as the primary tool for classes has provided an effective learning environment.
    \item A dedicated platform for P2P education would have enhanced my learning experience.
\end{enumerate}

\vspace{0.5em}
\textbf{Perspectives on Future Peer-to-Peer Education Initiatives}
\begin{enumerate}[label=\arabic*., nosep, itemsep=2pt, topsep=4pt, left=0pt]
    \setcounter{enumi}{8}
    \item I believe that expanding the peer-to-peer education model globally would help reduce educational inequalities.
    \item I would recommend P2P tutoring to other students:
    \begin{enumerate}[label*=\arabic*., nosep, itemsep=2pt, topsep=2pt, left=1em]
        \item During natural disasters
        \item In normal times, as a support tool alongside regular classroom education
    \end{enumerate}
    \item As a future university student, I would like to continue benefiting from P2P tutoring provided by senior or PhD students.
    \item I am eager to contribute to future P2P learning initiatives by volunteering as a tutor.
\end{enumerate}

\end{minipage}
\\
\multicolumn{2}{c}{
        \begin{minipage}{1.1\textwidth}
            \centering
            \includegraphics[width=\linewidth]{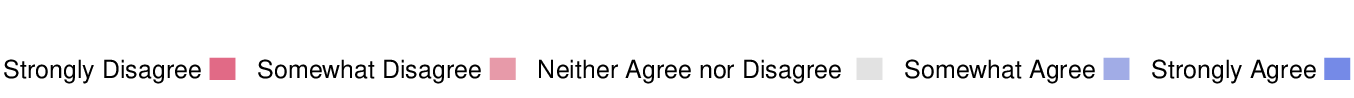}
            \captionof{figure}{Responses from volunteers (left) and tutoring receivers (right) in the Earthquake Solidarity Project.}
            \label{questionnaires}
        \end{minipage}
    } \\
\end{tabular}
\newpage
Regarding motivation, tutors prioritized intrinsic factors over financial compensation as their primary drivers. However, compensation was identified as a powerful lever for increasing potential time commitment: volunteers indicated a willingness to provide 2.6 hours/week unpaid (median: 2.0), but indicated readiness for an additional 3.4 paid hours (median: 2.0), an increase particularly pronounced among those who valued compensation more highly. Notably, the vast majority of learners expressed interest in becoming volunteer tutors themselves in the future.

Insights from open-ended responses provided a roadmap for future development. Tutors proposed creating a dynamic community where users could seamlessly switch between learner and educator roles, supported by AI-enhanced tutor-tutee pairings and opportunities for non-teaching volunteers to contribute resources. They also recommended leveraging advertising, social media, public announcements, and donor support to attract more participants. Learners, meanwhile, emphasized expanding the variety of subjects and increasing the availability of one-on-one classes.

\subsection{Visioning Session Insights: The Ideal P2P Ecosystem}
\subsubsection{\enquote{From Peers to the World}: Inclusivity, Access, and Safe Learning Spaces}
Across both groups, the envisioned platform is defined by a youth-led, inclusive, and socially purposeful learning ecosystem that scales beyond a local post-disaster setting. Tutors repeatedly framed the platform as internationally reachable —captured by the headline \enquote{From Peers to the World}—emphasizing global accessibility, equality, and cross-cultural connection. Learners echoed this expansionary vision through a visual metaphor: the \enquote{green globe,} which they interpreted as the platform's capacity to grow worldwide.

A second foundational value was \emph{peer-based approachability}, with the absence of hierarchical barriers central to this identity. Tutors argued that near-peer dynamics enable \enquote{comfort, accessibility, and openness in communication,} describing age proximity as an advantage that supports learning without triggering an immediate \enquote{authority figure} effect. Learners validated this, describing the environment as \enquote{non-authoritarian} and \enquote{sincere,} creating a climate where they could ask questions freely.

Third, the platform identity is explicitly relational and community-based. Learners prioritized psychological safety and belonging, describing the ideal environment not merely as a school but as a family-like community. This sentiment was vividly captured by Başak, who highlighted \enquote{a warm, sincere family atmosphere} characterized by \enquote{trust, love, and respect,} and by Ender, who reflected: \enquote{We weren’t just studying. We truly became like a family.} Sümeyye further highlighted \enquote{not being afraid to make mistakes} as a defining feature of this environment, while Ada described the platform as a \enquote{safe space} that helped learners feel understood and valued. Tutors echoed this emphasis on community building and trust as a central condition for sustaining motivation and participation over time, noting that \enquote{small connections turn into big impacts.}

Finally, participants from both groups framed the platform as a mechanism for opening doors. For learners, the \enquote{key} on their cover symbolized breaking prejudices, enabling access, and creating pathways forward. Tutors, in parallel, conceptualized the platform as an \enquote{awareness map} that makes educational need visible and provides a legitimate, structured entry point for those who want to help but \enquote{don't know where to start.}

\subsubsection{The Cycle of Intergenerational Solidarity}
A central element of the envisioned ecosystem was a self-renewing cycle of reciprocity, in which learners later return as tutors. Learners articulated this logic explicitly through the title \enquote{from generation to generation} and the headline \enquote{the peer-to-peer story continues}. Ender clarified that P2P was understood literally as \enquote{peer to peer}—learning first from more experienced near-peers, and then, once they reach a similar level of experience, passing forward that gift by \enquote{helping others the way they helped us}. Başak similarly described the model as iterative: tutors teach current learners, who then tutor younger learners in the future, sustaining the cycle.

This envisioned model of generational giving is not only aspirational but anchored in concrete role-model dynamics. Learners repeatedly positioned tutors as attainable exemplars—close enough in age and perspective to feel relatable, yet advanced enough to be motivating (e.g., studying or working in the university or profession the learner wants to reach). In this sense, tutoring is not only content support but also an informal pathway-orientation mechanism: learners learn how to progress by observing someone who has recently navigated similar transitions.

\subsubsection{Long-Term Sustainability}
While the cycle of solidarity addresses the long-term supply of human talent, tutors advocated for a pragmatic \emph{hybrid incentivization} model—a \enquote{blended space} that balances social impact with feasibility for university-student tutors—where educators can choose between volunteering for social impact and optional micro-earning opportunities for financial support. Şebnem emphasized the practical importance of this dual structure: \enquote{even a modest additional income} can function as a meaningful motivator because \enquote{money is a basic part of sustaining life,} especially given that the person providing education is often \enquote{also a student}. Büşra similarly envisioned \enquote{a platform where students can choose to volunteer if they want... but at the same time it also creates an environment where they can earn extra income if they wish.} In this sense, long-term sustainability is not positioned as a shift away from volunteering, but as a hybrid participation model that accommodates varying needs and life circumstances without eroding the platform’s solidarity-driven ethos.

Beyond financial incentives, sustainability was linked to professional growth. Tutors viewed the platform as a venue to \enquote{express themselves,} \enquote{gain confidence,} accumulate teaching experience, and build a professional network. This creates a value proposition where the volunteer role evolves from a one-way donor of time to a mutual beneficiary of skills, networking, and potential economic support.

In addition, tutors described the envisioned platform as a mechanism that reduces the friction of entry. By realizing the \enquote{awareness map} vision—making needs visible—the platform would create a clear pathway for potential volunteers. Within this structure, Emre suggested that involvement may begin informally and, over time, evolve into \enquote{something more professional} as the ecosystem grows, potentially bringing in more experienced contributors alongside near-peer tutors.

\subsubsection{Socio-Technical Barriers and Safeguarding Concerns}
Despite the optimistic vision, participants identified several hurdles to realization. Tutors, in particular, identified outreach and discoverability as the primary barrier to scaling. Şebnem observed: \enquote{The hardest part is reaching students and making them aware of the platform.} Elif concurred: \enquote{Awareness is the hardest part.} Emre reinforced this, noting that \enquote{spreading the idea is important} to reach those willing to help but lacking a clear entry point. This highlights the practical challenge inherent in the \enquote{awareness map} metaphor: visibility and dissemination are prerequisites for impact.

Tutors also raised concerns about quality control and safeguarding in an open, scaled ecosystem. Büşra noted the difficulty of \enquote{ensuring that everyone provides quality education,} while Emre highlighted risks of \enquote{misuse—people trying to exploit the system,} though he suggested that \enquote{a simple reporting system} could address this. İlker cautioned that personalization—one of the platform’s key promises—can elevate expectations. As he put it: \enquote{When something is personalized, people's expectations can rise a lot, and that can create problems.} Nefise added that \enquote{building and maintaining trust and sustaining motivation in the long term could be difficult.}

Learners reiterated technical and logistical barriers that shape participation in practice. Ender cited persistent connectivity issues. Başak noted the limitations of screen-based interaction: \enquote{Not being physically together makes it harder to read expressions.} Scheduling conflicts and digital fatigue also emerged as concerns, with Sümeyye mentioning \enquote{fatigue and lack of focus} alongside \enquote{time conflicts.} Finally, Ada noted that learners’ experiences differ, which can make it harder to \enquote{approach everyone in the same way} and to establish \enquote{a safe, trusting environment} consistently.

\section{Discussion}
\label{sec:discussion}
Interpreting our findings through several theoretical frameworks illuminates the interplay of individual, social, and technological factors that shape user engagement within online communities and can inform the design and sustainability of P2P learning platforms.

\subsection{Theory-Informed Analysis of a Volunteer-Driven P2P Ecosystem}
\paragraph{Self-Determination Theory (SDT) \cite{Deci_Ryan_2000}} provides a compelling narrative about the pivotal role of psychological needs in sustaining volunteer engagement within P2P learning ecosystems. Our findings show that intrinsic motivators dominate, directly satisfying all three of SDT's core needs: the desire for educational impact and witnessing tutee growth fulfills the need for \emph{Competence}; the emotional fulfillment from tutor-learner bonds and community belonging satisfies the need for \emph{Relatedness}; and the strong advocacy for flexible time commitment and micro-volunteering expresses the need for \emph{Autonomy}. Crucially, this demand for \emph{Autonomy} is not a desire for unbridled freedom but a practical need to align altruistic goals with real-life constraints.

By advocating for ad-hoc scheduling and urgency-responsive availability, tutors exemplify SDT's \emph{integrated regulation}, where actions harmonize intrinsic values with extrinsic demands (e.g., time scarcity). The visioning sessions deepened this insight. Tutors' advocacy for a hybrid incentivization model—allowing them to choose between volunteering and micro-earning—represents a further expression of \emph{Autonomy} that accommodates their financial reality. Rather than displacing intrinsic motivation, such optional incentives function as enablers that reduce practical barriers to sustained engagement. Additionally, tutors framed the platform as a venue for professional growth—to \enquote{express themselves,} \enquote{gain confidence,} and discover new teaching capabilities—further reinforcing the satisfaction of \emph{Competence}. Finally, their emphasis on \enquote{networking} and building \enquote{connections from very different places} extends the satisfaction of \emph{Relatedness} beyond the immediate tutor-learner bond to a broader global community.

These insights underscore the need for P2P platforms to prioritize features that holistically support volunteers by satisfying their needs for \emph{Autonomy} (through flexibility and choice), \emph{Competence} (by highlighting impact and enabling professional growth), and \emph{Relatedness} (by fostering community and cross-cultural networking), thereby sustaining engagement.

\paragraph{Social Identity Theory (SIT) \cite{Tajfel_Turner_2004}} sheds light on the nuanced role of age proximity in sculpting tutor–learner dynamics. Learners’ preference for tutors close in age underscores SIT’s premise that a shared \emph{in-group} identity, based on age, reduces hierarchical distance, fostering openness, trust, and communication. The visioning sessions revealed that this \emph{in-group} bond is cemented by a deeper, superordinate identity rooted in shared lived experience, specifically the collective trauma of the earthquake, which created a profound sense of solidarity.
 
Yet, tutors' apprehensions about blurred authority boundaries highlight a core SIT conflict: this shared identity simultaneously weakens the \emph{out-group} distinction required for a traditional instructor role. This creates a tension between camaraderie and guidance—a challenge created by the tutor simultaneously occupying both \emph{in-group} (peer) and \emph{out-group} (authority) roles. 

However, the \enquote{Generation to Generation} dynamic suggests a resolution to this tension: learners view tutors as aspirational \enquote{role models} who represent their ideal future selves. In SIT terms, tutors serve as the \emph{prototypical members} of the group to which learners aspire to belong. This grants them a form of referent authority that commands respect without requiring rigid hierarchy. This suggests that P2P platform design should reinforce tutors’ professional identity (e.g., through badges) to maintain authority while leveraging the benefits of this distinct \emph{in-group} rapport.

\paragraph{Social Capital Theory (SCT) \cite{Putnam_1993}} illuminates how relational resources sustain engagement in P2P learning. Tutors’ sense of belonging to a community that \enquote{outlives its individual members} and learners’ comfort in smaller, personalized groups reflect the accumulation of \emph{bonding social capital} (the strong, trusting ties within a group), which fosters trust and reciprocity. The visioning sessions underscored the depth of these ties, with learners explicitly characterizing the environment as a \enquote{family-like} \enquote{safe space}—indicating that bonding capital here serves as a resource for psychological recovery and continued participation.

Additionally, tutors’ aspiration for professional development indicates the creation of \emph{bridging social capital} (links to external resources and networks), as the platform allows volunteers to convert their participation into tangible professional assets, including an enhanced CV and valuable references. In SCT terms, these bridges expand access to information, opportunities, and recognition that are not available within participants’ immediate circles.

Both groups extended this view by envisioning the platform's potential for global reach and cross-cultural connection. Tutors captured this aspiration in their headline \enquote{From Peers to the World,} while learners echoed it through the visual metaphor of a \enquote{green globe}—symbolizing the platform's capacity to scale bridging capital beyond local networks to an international level. However, shifting from local, high-trust contexts to a global scale highlights the need for platforms to incorporate trust-building mechanisms (e.g., reputation systems, verification, and transparent reporting) to intentionally cultivate these weak ties, thereby sustaining participation.

Perhaps most significant is the cycle of intergenerational reciprocity envisioned by learners, wherein current tutees aspire to return as future tutors. From an SCT perspective, this cycle can be interpreted as a mechanism of social-capital reproduction: rather than merely accumulating or exchanging capital, the ecosystem is designed to reproduce it across generations, converting today's beneficiaries into tomorrow's benefactors.

\paragraph{Media Richness Theory (MRT) \cite{Daft_Lengel_1986}} helps interpret the impact of communication modalities on engagement. In line with MRT’s principle that complex, equivocal tasks like teaching require richer media, our findings reveal a clear demand for a richer communication environment, evidenced by tutors’ advocacy for interactive elements and learners’ appreciation for immediate feedback. The visioning sessions reinforced this need: learners identified the inherent constraints of screen-based interaction, specifically the inability to \enquote{read expressions}. However, participants also reported achieving rich relational outcomes through this ostensibly leaner medium, suggesting that tutors' deliberate integration of informal elements—such as games, casual conversations, and personalized attention—compensated for technological limitations by augmenting the medium's effective richness through pedagogical design. This suggests that platform designers should prioritize integrating richer media features while also scaffolding interaction practices that amplify the perceived richness of online environments.

\subsection{Theoretical Implications} 
This study’s analysis of how a grassroots P2P initiative evolved from crisis response to an enduring socio-technical infrastructure yields key theoretical implications at the intersection of digital volunteering and peer learning. First, extending Self-Determination Theory to ``crisis-catalyzed volunteerism,'' we find that acute situational urgency can intensify intrinsic motivations such as impact and solidarity, overriding barriers like time scarcity and facilitating sustained engagement. The visioning sessions further refine this by suggesting that as urgency fades, motivation must transition from purely altruistic drives to \emph{hybridized autonomy}—where flexible scheduling, micro-volunteering options, and the choice to combine volunteering with micro-earning collectively empower participants to align their altruistic values with evolving personal and economic realities, thereby reducing dropout risk and preventing burnout.

Second, we add nuance to Social Identity Theory by identifying a ``dual-identity paradox'': the tutor's shared \emph{in-group} identity with learners fosters trust but weakens the \emph{out-group} distinction required for authority. To explain how effective tutors navigate this, we introduce the concept of ``pedagogical code-switching'': the ability to strategically shift between a peer identity (to build rapport) and an instructor identity (to guide learning). We propose that this capacity is a critical, yet previously unarticulated, skill for effective classroom management in non-reciprocal, far-peer tutoring.

Furthermore the visioning sessions revealed an additional identity mechanism: aspirational role modeling. Learners positioned tutors not merely as peers but as attainable exemplars of their desired future selves—individuals already inhabiting the university or profession the learner aspires to reach. This temporal dimension extends SIT by showing that in-group identity can be simultaneously present and aspirational (the group the learner strives to join in future).

Third, extending Social Capital Theory, we identify a novel mechanism of social capital reproduction through what we term the \emph{cycle of intergenerational reciprocity}—a gift-economy dynamic in which today's beneficiaries become tomorrow's benefactors. Learners explicitly articulated their intention to return as tutors transforming the ecosystem from a capital-consuming initiative into a self-regenerating network. This extends SCT beyond the accumulation and exchange of bonding and bridging capital to explain how volunteer-driven ecosystems can achieve long-term sustainability.

Fourth, we challenge strict interpretations of Media Richness Theory by demonstrating that \emph{pedagogical richness}—the deliberate infusion of socio-emotional cues through interaction routines—can compensate for technical leanness. This implies that perceived richness is not solely a property of the medium but can be socially constructed through interaction design. Therefore, effective P2P design must couple richer affordances (e.g., interactive whiteboards) with structured protocols that amplify social presence, ensuring that feedback and emotional signals remain visible even under bandwidth constraints.

Finally, by tracing how an improvised online learning environment morphed into a long-term learning infrastructure, we identify the socio-technical mechanisms that convert spontaneous volunteer energy into resilient educational ecosystems. Collectively, these insights provide a transferable analytic lens for scholars examining future crises that catalyze digitally mediated learning communities.

\subsection{Design Principles for Sustainable P2P Learning Ecosystems} 
Our findings offer a blueprint for designing and sustaining volunteer-driven P2P educational ecosystems that expand equitable access to quality learning. We propose five key design principles:

\textbf{Design for Flexible yet Continuous Engagement.} P2P platforms should offer a hybrid model that combines stable, long-term group sessions with flexible micro-volunteering options for one-on-one support. This approach maximizes both tutor participation and learner continuity. To streamline coordination, platforms should integrate automated scheduling and communication tools. Additionally, platforms should offer a hybrid incentivization model that allows tutors to choose between pure volunteering and optional micro-earning opportunities based on their circumstances—accommodating the reality that tutors are often students themselves with financial constraints.

\textbf{Build a Diversified Resource Strategy.} Sustainable operations depend on diversified resource streams. This includes securing financial support through sponsorships, grants, or corporate partnerships, as well as forging strategic collaborations with educational institutions and NGOs to gain access to vital resources, expand recruitment pools, and broaden community reach. Critically, since platform awareness and outreach represent a primary barrier to scaling, resource strategies must explicitly allocate effort toward discoverability and dissemination—not only volunteer recruitment.

\textbf{Foster a Mutually Beneficial Volunteer Ecosystem.} To sustain volunteer engagement, programs should offer a blend of professional development, formal recognition (e.g., certificates, digital badges), and community-building opportunities. Rather than relying solely on altruism, the ecosystem should function as a venue for professional growth and networking. Integrating anonymous, multi-modal feedback channels into the platform can further support ongoing improvement and foster a sense of belonging.

\textbf{Enable Data-Informed Personalization.} Platforms should leverage personalization as a core advantage of the P2P model. This can be achieved by using brief pre-assessments to inform matching algorithms that create compatible and effective tutor-learner pairings. Beyond academic needs, these algorithms should also consider aspirational fit—pairing learners with tutors who represent their desired future trajectories—to leverage the motivational power of role modeling.

\textbf{Design for Intergenerational Continuity.}  Sustainable P2P ecosystems should treat role transition (learner → tutor) as a first-class design goal by building an explicit learner-to-tutor pathway. This fosters a culture of reciprocity where learners are inspired to become future role models, transforming the ecosystem into a regenerative community that passes knowledge and solidarity from one generation to the next.

Together, these design principles operationalize a vision for solidarity-based learning, guiding the creation of scalable learning ecosystems that advance educational equity.

\section{Conclusions}
\label{sec:conclusion}
This study reports on a crisis-born initiative that transcended its emergency origins to become an enduring infrastructure of solidarity. By triangulating retrospective operational data with prospective visioning workshops, we moved beyond a descriptive account of crisis response to conceptualize the mechanisms underlying a sustainable P2P learning ecosystem and its pathways to scaling.

Our findings yield several contributions. Theoretically, we extend Self-Determination Theory to \emph{crisis-catalyzed volunteerism,} demonstrating that acute situational urgency can intensify intrinsic motivations—such as impact and solidarity—overriding barriers like time scarcity. We further refine this by introducing the concept of \emph{hybridized autonomy}—wherein, as urgency fades, flexible time commitment, micro-volunteering options, and optional micro-earning opportunities collectively empower participants to align altruistic values with evolving personal and economic realities, thereby sustaining engagement. We enrich Social Identity Theory by identifying a \emph{dual-identity paradox} and proposing \emph{pedagogical code-switching} as a mechanism through which tutors navigate the tension between peer rapport and instructional authority. The visioning sessions revealed an additional identity mechanism—\emph{aspirational role modeling}—extending SIT by showing that in-group identity can be simultaneously present and aspirational. We also extend Social Capital Theory by conceptualizing the \emph{cycle of intergenerational solidarity} as a mechanism of social capital reproduction. Finally, we challenge strict interpretations of Media Richness Theory by demonstrating that \emph{pedagogical richness} can compensate for the technical leanness of digital media.

Practically, we distill five design principles for sustainable P2P learning ecosystems. These principles offer actionable guidance for practitioners seeking to scale similar 
initiatives while preserving the human-centered qualities—trust, solidarity, and 
relational depth—that distinguished this case.

Looking ahead, the strategic integration of Generative AI (GenAI) represents the next frontier for such ecosystems. The central challenge lies in designing human-AI configurations that amplify, rather than displace, human connection. Future inquiries should explore AI \enquote{co-pilots} that alleviate volunteer time constraints by generating personalized pedagogical resources, alongside AI-tutors that provide persistent scaffolding for learners between human mentorship sessions. Designing and evaluating these symbiotic partnerships will be essential for creating sustainable and equitable learning systems in the \enquote{post-COVID-19 and generative AI era}~\cite{Marcinkowski_etal_2025}.

\section*{Acknowledgements}
A. Przybylek’s contribution was funded by Taighde Éireann – Research Ireland (Grant No.13/RC/2094\_2). Co-funded by the European Union (SyMeCo, Grant No.101081459). The views expressed are solely those of the author(s). Neither the European Union nor the European Research Executive Agency can be held responsible for them.
%
%
%
\bibliographystyle{splncs04}
\bibliography{references}

\end{document}